\documentclass[aip,jcp,reprint,amsmath,amssymb,floatfix,citeautoscript]{revtex4-2}
\usepackage{cancel}
\usepackage{amsmath}
\usepackage{mathtools}
\usepackage{physics}
\usepackage{graphicx}
\usepackage{amssymb}
\usepackage{amsthm}
\usepackage{bm}
\usepackage{dcolumn}
\usepackage{braket}
\usepackage{ragged2e}
\usepackage{txfonts}
\usepackage[version=3]{mhchem}
\usepackage[T1]{fontenc}
\usepackage[colorlinks=true,allcolors=blue]{hyperref}
\usepackage[capitalise]{cleveref}

\newcommand{\vq}{{\bm{q}}}
\newcommand{\vn}{{\bm{n}}}
\newcommand{\vm}{{\bm{m}}}
\newcommand{\vp}{{\bm{p}}}
\newcommand{\vA}{{\bm{a}}}

\allowdisplaybreaks

\begin{document}
\title{Vibrational heat-bath configuration interaction with semistochastic perturbation theory 
using harmonic oscillator or VSCF modals}

\author{Henry K. Tran}
\affiliation{Department of Chemistry, Columbia University, New York, New York 10027, USA}
\author{Timothy C. Berkelbach}
\email{t.berkelbach@columbia.edu}
\affiliation{Department of Chemistry, Columbia University, New York, New York 10027, USA}

\begin{abstract}
Vibrational heat-bath configuration interaction (VHCI)---a selected configuration
interaction technique for vibrational structure theory---has recently been
developed in two independent works [J.\ Chem.\ Phys.\ \textbf{154}, 074104 (2021);
Mol.\ Phys.\ \textbf{119}, e1936250 (2021)], where it was shown to provide
accuracy on par with the most accurate vibrational structure methods with a low
computational cost.  Here, we eliminate the memory bottleneck of the
second-order perturbation theory (PT2) correction using the same
(semi)stochastic approach developed previously for electronic structure theory.
This allows us to treat, in an unbiased manner, much larger perturbative
spaces, which are necessary for high accuracy in large systems.  Stochastic
errors are easily controlled to be less than 1~cm$^{-1}$.  We also report two
other developments: (i) we propose a new heat-bath criterion and an associated
exact implicit sorting algorithm for potential energy surfaces expressible as a
sum of products of one-dimensional potentials; (ii) we formulate VHCI to use a
vibrational self-consistent field (VSCF) reference, as opposed to the harmonic
oscillator reference configuration used in previous reports. Interestingly, we
find that with VSCF, the minor improvements to the accuracy are outweighed by
the much higher computational cost associated with the loss of sparsity in the
Hamiltonian and integrals transformations needed for matrix element evaluation.
\end{abstract}

\maketitle

\section{Introduction}
\label{sec:introduction}
Vibrational spectroscopy has long been a tool for chemical fingerprinting and 
analysis~\cite{vib-apps-1}, and experimental advances have enabled high resolution 
infrared (IR) and near-infrared (NIR) spectra of large biological and material 
systems~\cite{vib-bio-1, vib-materials-2}. 
To achieve comparably high resolution, computational methods must be developed that accurately
treat vibrational anharmonicity.
The most affordable computational methods are mean-field theory, commonly known as vibrational self-consistent field (VSCF) theory~\cite{vscf-1, vscf-2, handy-vib-1982, handy-rovib-1986, vscf-11, vscf-10, vscf-9, vscf-12, vscf-3, vscf-4, vscf-5, vscf-6, vscf-7, vscf-8}, or bare perturbation theories~\cite{vvpt-1, vvpt-2, vpt2-1, vpt2-2, vpt2-3, vpt2-3, vpt2-4,vpt2-5, vpt2-6, vpt2-7, vpt4-1, vpt4-2, vpt4-3}.
More accurate results can be obtained by improving upon a VSCF reference via perturbation theory~\cite{vscf-mp2-1, vscf-mp2-2}, configuration interaction~\cite{shavitt, vscf-8, vscf-ci-1, vscf-ci-2}, or coupled-cluster theory~\cite{bartlett-2007, vcc-1, vcc-2, vcc-3}, paralleling methodology developments in electronic structure theory.

More modern electronic structure theory methods have also been adapted for vibrational structure, including vibrational density-matrix renormalization group~\cite{vdmrg} and vibrational selected configuration interaction methods, the latter of which prunes the exponentially large configuration space.
Here, we continue the development of vibrational heat-bath configuration interaction (VHCI), which is based on an analogous method
in electronic structure theory~\cite{hci-1, hci-2, hci-3, hci-4} and which was developed independently in Refs.~\onlinecite{vhci-1} and \onlinecite{vhci-2}.
These initial implementations of VHCI, combined with second-order Epstein-Nesbet perturbative corrections (VHCI+PT2), demonstrated sub-wavenumber accuracy for hundreds of vibrational states on molecules containing tens of degrees of freedom.

Despite promising preliminary results, VHCI has room for improvement. 
First, the PT2 correction is responsible for a severe memory bottleneck, which precludes large perturbative spaces but which has been eliminated in electronic HCI through a stochastic method~\cite{hci-2}.
Second, VHCI has only been performed using configurations of harmonic oscillator eigenfunctions as opposed to vibrational self-consistent field (VSCF) modals~\cite{vscf-1, vscf-2}, which partially capture anharmonicity at the mean-field level and are commonly used in vibrational structure methods~\cite{vscf-ci-1, vscf-ci-2, vcc-1, vcc-2, vcc-3}. 
Third, VHCI benefits through implicit sorting of Hamiltonian matrix elements and explicitly generating only the largest matrix elements; previous studies either approximated the elements in the implicit sort~\cite{vhci-1} or explicitly generated all matrix elements for the sorting~\cite{vhci-2}.

In this paper, all three of these directions are explored and their effects are benchmarked. 
We modify VHCI to use VSCF configurations, finding more accurate results, but with a steeper computational cost. 
A new heat-bath criterion is presented that implicitly sorts exact matrix elements at no cost to computational time and a negligible memory cost for potential energy surfaces that are expressible as a sum of products of one-dimensional potentials.
Finally, and most significantly, VHCI with (semi)stochastic PT2 corrections is implemented, demonstrating the same accuracy as VHCI+PT2 with a much lower computational and memory cost.

\section{Theory}
\label{sec:theory}

The vibrational Hamiltonian for a molecule with $N=3N_\mathrm{atom}-6$ vibrational degrees of freedom is
\begin{equation}
\begin{split}
    H =
    \sum_{i=1}^N \left[ -\frac{1}{2}\frac{\partial^2}{\partial q_i^2} \right]
    + V(\vq),
\end{split}
\end{equation}
where $q_i$ are mass-weighted normal mode coordinates and $V(\vq)$ is the
$N$-dimensional Born-Oppenheimer potential energy surface (PES). 
The representation of the PES is critical in numerical calculations.
An especially convenient representation is as a sum of products (SOP) of one-dimensional
potentials (sometimes called single-particle potentials),
\begin{equation} \label{eq:product_potential}
V(\vq) = \sum_{p_1=1}^{P_1}\cdots\sum_{p_N=1}^{P_N} V_{p_1\cdots p_N} \prod_{i=1}^N v^{(i,p_i)}(q_{i}),
\end{equation}
which yields the simple evaluation of matrix elements between product states.
Specifically, letting 
\begin{equation} \label{eq:vibwf}
    \Phi_\vn(\vq) = \prod_{i=1}^{N} \phi^{(i)}_{n_i}(q_i)
\end{equation}
where $\vn = (n_1, n_2, \ldots)$ and $n_i$ is the excitation level
of mode $i$, PES matrix elements are given by
\begin{equation}
\bra{\Phi_\vm} V \ket{\Phi_\vn} = \sum_{p_1\cdots p_N} V_{p_1\cdots p_N} \prod_{i=1}^N v^{(i,p_i)}_{m_in_i},
\end{equation}
where $v^{(i,p_i)}_{m_in_i} \equiv \bra{\phi_{m_i}^{(i)}} v^{(i,p_i)} \ket{\phi_{n_{i}}^{(i)}}$.
Thus, an $N$-dimensional integral is approximated by a sum of products of $N$ one-dimensional integrals. 
This PES representation has been especially common in calculations with the multi-configurational
time-dependent Hartree approach~\cite{meyer-2008, carrington-2011, carrington-2016}.
We note that the familiar Taylor series expansion of the PES is a limiting
case of the general SOP potential. Another powerful representation of the PES is the $n$-mode 
representation~\cite{vscf-1, vscf-2}, although this is harder to use in VHCI due to the large number of
unique many-body integrals it generates. We consider this an important goal for future work.

\subsection{Vibrational SCF} \label{sec:theory-vscf}

Previous work on VHCI~\cite{vhci-1,vhci-2} used product-state configurations built from harmonic oscillator (HO) eigenfunctions.
Here, we additionally explore the use of modals generated by the vibrational self-consistent field (VSCF) 
procedure~\cite{vscf-1,vscf-2, vscf-3}, which we briefly review to establish notation.
The modals $\phi^{(i)}_{n_i}(q_i)$ in Eq.~\eqref{eq:vibwf} are obtained variationally
as the eigenfunctions of a one-dimensional Hamiltonian operator,
\begin{equation}
    \hat h^{(i)} \phi^{(i)}_{n_i}(q_i) 
    = \left[-\frac{1}{2}\frac{\partial^2}{\partial q_i^2}+w_{\vn}^{(i)}(q_i)\right]\phi^{(i)}_{n_i}(q_i)
    = \varepsilon^{(i)}_{n_i} \phi^{(i)}_{n_i}(q_i)
\end{equation}
where the effective potential for mode $i$ is
\begin{equation}
\begin{split}
    w_{\vn}^{(i)}(q_i) &= \left\langle \prod_{j\neq i} \phi^{(j)}_{n_j}\Big| V \Big| 
    \prod_{j\neq i} \phi^{(j)}_{n_j} \right\rangle
\end{split}
\end{equation}
The modals can be calculated on a grid or can be expanded in some basis set,
\begin{equation}
\phi^{(i)}_{n_i}(q_i) = \sum_{\mu=1}^{K} C_{\mu n_i}^{(i)} \chi^{(i)}_{\mu} (q_i),
\end{equation}
where $K$ is the number of basis functions per mode.  Here, we use a basis set
expansion and we consider only modals optimized for the ground state with
$\vn=(0,0,\ldots)$, although state-specific or state-averaged calculations
could be considered.

The modal coefficients $C^{(i)}_{\mu n_i}$
are obtained self-consistently as the eigenvectors of a one-body Hamiltonian matrix
\begin{align}
    \mathbf{h}^{(i)}\mathbf{C}^{(i)}_{n_i} &= \varepsilon_{n_i}^{(i)} \mathbf{C}^{(i)}_{n_i}, \\
\label{eq:fock}
    h^{(i)}_{\mu\nu} &= t^{(i)}_{\mu\nu} + w^{(i)}_{\mu\nu},
\end{align}
where $t^{(i)}_{\mu\nu} = -(1/2)\langle \chi_\mu^{(i)}|\partial^2/\partial q_i^2|\chi_\nu^{(i)}\rangle$
and  $w^{(i)}_{\mu\nu}$ includes a mean-field potential due to the other modes.
With the SOP representation of the PES in Eq.~\eqref{eq:product_potential},
the fundamental ingredients are the basis function integrals over the
one-dimensional potentials $v^{(i,p_i)}_{\mu\nu}$, 
of which there are $\mathcal{O}(P N K^2)$, where $P$ is a typical value of $P_i$.
The integrals can be transformed to the modal basis,
\begin{equation} \label{eq:Ytensor}
v^{(i,p_i)}_{m_in_i} = \sum_{\mu\nu}C_{\mu m_i}^{(i)} C_{\nu n_i}^{(i)} v^{(i,p_i)}_{\mu\nu},
\end{equation}
with a CPU cost of $\mathcal{O}(P N K^3)$ for all modes.
The effective potential for ground-state VSCF is then given by
\begin{subequations} \label{eq:vscf_pot_prod}
\begin{align}
w_{\mu\nu}^{(i)}
&= \sum_{p_i} v^{(i,p_i)}_{\mu\nu} \bar{v}^{(i,p_i)}, \\
\bar{v}^{(i,p_i)} &= \sum_{p_1\cdots p_N}^\prime
    V_{p_1\cdots p_N} \prod_{j\neq i}^{N} v^{(j,p_j)}_{00}
\end{align}
\end{subequations} 
where the primed summation excludes $p_i$ and $v_{00}$ is understood to be
in the modal basis.  Evaluating the above sum for all
modes costs $\mathcal{O}(P^{N} N^2)$ CPU operations.

Each iteration of the VSCF cycle requires only the lowest-energy eigenvector
($n_i=0$) of the effective Hamiltonian matrix in Eq.~\eqref{eq:fock} for each
mode.  At convergence, full diagonalization of the Hamiltonian matrix
additionally defines the $(K-1)$ excited-state modals ($n_i > 0$) to be used in
a post-VSCF treatment.

\subsection{Vibrational CI} \label{sec:theory-vci}

In VCI, we variationally calculate multiconfigurational wavefunctions
\begin{equation}
    |\Psi\rangle = \sum_{\vn\in \mathcal{V}} c_\vn|\vn\rangle,
\end{equation}
where $c_\vn$ are eigenvectors of the Hamiltonian matrix in a basis of 
configurations $|\vn\rangle$ belonging to the variational space $\mathcal{V}$,
and $\langle \vq|\vn\rangle \equiv \Phi_{\vn}(\vq)$.
Using a SOP potential, the Hamiltonian matrix elements are
\begin{equation} \label{eq:ham_prod}
\langle \vm | H | \vn \rangle
= \sum_i t^{(i)}_{m_i n_i} \prod_{j\neq i} \delta_{m_j n_j}
    + \sum_{p_1\cdots p_N} V_{p_1\cdots p_N} \prod_i v^{(i,p_i)}_{m_i n_i}.
\end{equation}

\subsection{Vibrational heat-bath CI with VSCF reference} \label{sec:theory-vhci}

In conventional VCI calculations, the variational space $\mathcal{V}$ commonly
contains all configurations satisfying certain excitation restrictions.
In contrast, selected CI methods aim to iteratively build the space $\mathcal{V}$
to include only the most important configurations. In the heat-bath CI algorithm,
which was originally developed for electronic structure calculations~\cite{hci-1,hci-2,hci-3,hci-4,hci-6,hci-7},
configurations are added to $\mathcal{V}$ on the basis of the current
variational wavefunction and the magnitude of the Hamiltonian matrix elements connecting
configurations in $\mathcal{V}$ to those outside of $\mathcal{V}$.
Specifically, a new configuration $\vm$ is added if
\begin{equation}
\label{eq:vhci_crit}
    |H_{\vm\vn} c_\vn| > \epsilon_1
\end{equation}
where $\epsilon_1$ is a user-defined threshold. The power in this criterion lies
in the observation that the elements belonging to the columns of the Hamiltonian
matrix can be implicitly sorted, yielding an efficient algorithm for the identification of
new configurations to add. Importantly, unimportant configurations are never considered.

In electronic structure theory---because of the two-body interaction, the
quantum statistics of electrons, and the associated exclusion principle---the
essential Hamiltonian matrix elements depend only on the identity of the four
orbitals involved in a two-orbital replacement, $\langle ab||ij\rangle$. 
However, the matrix elements in VCI are significantly more complicated.
The scaling of the number of nonzero elements is more steep, due to the $n$-body 
interaction. Moreover, the matrix elements carry a dependence on both the
\textit{identity} and the \textit{excitation level} of the participating modes,
due to the lack of an exclusion principle.
Specifically, if the PES representation explicitly couples up to $n$
modes at a time (where $n$ is commonly 2--6, but can be as large as $N$), 
then the number of unique
elements in each column of $H$ is $\mathcal{O}(N^n K^n)$, which is too large to
store and sort. 
Past works have either applied an approximate heat-bath
criterion~\cite{vhci-1} or studied molecules whose size and symmetry
yield a sufficiently sparse Hamiltonian matrix to allow for
explicit storage and sorting~\cite{vhci-2}.
Moreover, these previous works used HO modals in constructing
configurations, and it is unclear to what extent their approximations or
restrictions can be maintained with VSCF modals (for which the Hamiltonian is
far less sparse).  In this work, we present a new, ``exact'' heat-bath
criterion and associated sorting algorithm for the case of a SOP potential.
Importantly, our method is applicable to both HO and VSCF modals.

The basic observation behind the implementation of an exact and efficient heat-bath search
is that the many unique Hamiltonian matrix elements are formed from products of 
the far fewer $v^{(i,p_i)}_{n_i m_i}$, i.e., the matrix elements of the one-dimensional
potentials in the SOP representation; as already mentioned, there are only $\mathcal{O}(PNK^2)$
such elements.
Following the heat-bath algorithm in previous work~\cite{vhci-1}, 
we test configurations according to the slightly weaker condition,
\begin{align}
\label{eq:hb}
    \left|V^{(\vp)}_{\vm \vn} c_\vn\right| &> \epsilon_1 \\
    V^{(\vp)}_{\vm \vn} &= V_{\vp} \prod_i v_{m_i n_i}^{(i,p_i)}
\end{align}
where $V(\vq) = \sum_{\vp} V^{(\vp)}(\vq)$ introduces shorthand notation
for Eq.~\eqref{eq:product_potential} and $\vp = (p_1, \ldots, p_N)$.
We test Eq.~(\ref{eq:hb}) for all $\vp$ defining the SOP potential.

With these observations, the essential VHCI trick is to pre-sort the columns
of the matrices $v^{(i,p_i)}_{m_i n_i}$ in descending order by magnitude.
Then, for a given configuration $|\vn\rangle$ (with coefficient $c_\vn$) and term in the SOP potential $\vp$,
it is straightforward 
to descend the list of many-body matrix elements $V^{(\vp)}_{\vm\vn}$
according to their magnitude, without checking almost all of the configurations that
will not satisfy Eq.~(\ref{eq:hb}).
Specifically, we use the (sorted) $n_i$th column of each matrix $v^{(i,p_i)}_{m_i n_i}$;
the product of the first (largest) element of each column is responsible for the largest
matrix element $V^{(\vp)}_{\vm\vn}$. If Eq.~(\ref{eq:hb}) is satisfied, we add the configuration
$|\vn\rangle$ defined by the row index $n_i$ of the sorted columns. 
By inspecting the second element of each sorted column, the
next-largest many-body matrix element $V^{(\vp)}_{\vm\vn}$ can be identified. This process can
be repeated until passing below the threshold in Eq.~(\ref{eq:hb}).

To obtain multiple vibrational states, this procedure can be performed for each one,
\begin{equation}
	\ket{\Psi_v} = \sum_{\vn \in \mathcal{V}} c_{\vn v} \ket{\vn}.
\end{equation}
However, if all states are sought at once, then this is equivalent to
performing the heat-bath selection with the single vector 
$c^\text{max}_\vn \equiv \max_v c_{\vn v}$.  In this case, the variational
space can quickly grow to be quite large, however its growth is sublinear in
the number of desired states because many important configurations are shared.

\subsection{Semistochastic VPT2} \label{sec:theory-sspt2}
The heat-bath approach can also be used with Epstein-Nesbet perturbation theory
to produce a state-specific energy correction,
\begin{equation} \label{eq:pt2}
\Delta E^{(2)}_v(\epsilon_2) \approx \sum_{\vA \in \mathcal{P}_v(\epsilon_2)} 
    \frac{\left(\sum_{\vn \in \mathcal{V}} H_{\vA\vn} c_{\vn v}\right)^2}{E_v - H_{\vA\vA}},
\end{equation}
where $E_v$ is the variational energy of the state defined by $c_{\vn v}$ and
$\mathcal{P}_v(\epsilon_2)$ is the perturbative space of configurations
$|\vA\rangle$ satisfying the heat-bath criterion in Eq.~\eqref{eq:hb} (with
$\epsilon_2 < \epsilon_1$) that are not already in $\mathcal{V}$, which can be
identified efficiently in the same manner as described above.  The Hamiltonian
matrix elements are calculated from the procedure described in
Sec.~\ref{sec:theory-vci}.  The most CPU-efficient implementation of
Eq.~\eqref{eq:pt2} requires a list that is the size of the perturbative space,
where each element stores a running sum of contributions from configurations
$|\vn\rangle$.  For large variational spaces, the size of the perturbative
space can become enormous for reasonable values of $\epsilon_2$, and the
storage of this list becomes a memory bottleneck.

Electronic HCI has addressed this memory problem with a stochastic formulation
of the PT2 correction\cite{hci-2,hci-3,hci-4,hci-7}, which we apply to the
vibrational problem in this work. We briefly review stochastic PT2 (SPT2) and
semistochastic PT2 (SSPT2).  In SPT2, we sample $N_c$ configurations from the
variational space with probability
\begin{equation}
    \pi_\vn = \frac{|c_{\vn v}|}{\sum_{\vm \in \mathcal{V}} |c_{\vm v}|}
\end{equation}
yielding an integer-valued population $w_\vn$ for each configuration (note that
configurations can be selected more than once, i.e., $w_\vn$ can be greater
than 1 and the number of unique configurations sampled can be less than $N_c$).  
Let $\mathcal{V}^s$ be the space of sampled variational
configurations with non-zero population.  The perturbative space
$\mathcal{P}_v^s(\epsilon_2)$ is determined by only the $O(N_c)$ configurations in
$\mathcal{V}^s$ (and their corresponding coefficients $c_{\vn v}$).  The
perturbative correction can be calculated from the unbiased expectation,
\begin{equation} \label{eq:spt2}
\begin{gathered}
    \Delta E_v^{(s2)}(\epsilon_2) = \frac{1}{N_c(N_c-1)} \Bigg\langle \sum_{\vA \in \mathcal{P}_v^s(\epsilon_2)} \left[ \left( \sum_{\vn\in\mathcal{V}^s} \frac{w_\vn c_{\vn v} H_{\vA\vn}}{\pi_\vn} \right)^2 \right. \\
    \left. + \sum_{\vn\in\mathcal{V}^s}\left( \frac{w_\vn(N_c - 1)}{\pi_\vn} - \frac{w_\vn^2}{\pi_\vn^2} \right) c_{\vn v}^2 H_{\vA\vn}^2 \right] \frac{1}{E_v - H_{\vA\vA}} \Bigg\rangle,
\end{gathered}
\end{equation}
where the term in brackets is averaged over $N_s$ independent samples.  For each
independent realization, the variational space is resampled and the reduced
perturbative space is redetermined.  For sufficiently small $N_c$, the size of
perturbative space $\mathcal{P}_v^s(\epsilon_2)$ is much smaller than that
without sampling $\mathcal{P}_v(\epsilon_2)$; thus the memory bottleneck can be
eliminated at the expense of embarrassingly parallel sampling.  However, this
stochastic approach can have a large statistical error unless $N_c$ and/or
$N_s$ are sufficiently large. 

To offset the statistical error, part of the perturbative space can be treated
deterministically and only the remainder treated stochastically, leading to the
semistochastic correction 
\begin{equation}
\Delta E_v^{(ss2)}(\epsilon_2^d, \epsilon_2) = 
\Delta E_v^{(2)}(\epsilon_2^d) 
+ \left[ \Delta E_v^{(s2)}(\epsilon_2) - \Delta E_v^{(s2)}(\epsilon_2^d) \right] 
\end{equation}
where $\epsilon_2 < \epsilon_2^d$ defines the stochastic part of the
perturbative space and $\epsilon_2^d$ defines the deterministic part. In order
to reduce the statistical noise, correlated sampling is used: each time the
variational space is sampled, both $\mathcal{P}_v^s(\epsilon_2)$ and
$\mathcal{P}_v^s(\epsilon_2^d)$ are generated from the same sample in order to
calculate the stochastic corrections.

\section{Results} \label{sec:results}

\subsection{Computational Details}

For consistency with previous works, where VHCI has been previously applied and where
high-accuracy benchmark results exist, we use an expansion of the potential
energy surface in powers of normal-mode coordinates, 
\begin{equation} \label{eq:taylor_potential}
\begin{split}
        V(\vq) &= \sum_i V_{ii} q_i^2 + \sum_{i\le j\le k} V_{ijk} q_i q_j q_k + \ldots \\
        &= \sum_i W_{ii} \bar{q}_i^2 + \sum_{i\le j\le k} W_{ijk} \bar{q}_i \bar{q}_j \bar{q}_k + \ldots
\end{split}
\end{equation}
where $W_{ijk\ldots} = V_{ijk\ldots}/(2^p \omega_i \omega_j \omega_k \ldots)^{1/2}$,
$p$ is the order of the expansion, $\omega_i$ is the normal-mode frequency of
the $i$\textsuperscript{th} mode, and $\bar{q}_i = (a^\dagger_i + a_i)$. 
As already mentioned, the normal-mode expansion~(\ref{eq:taylor_potential}) is a
special case of the SOP potential~(\ref{eq:product_potential}), with
\begin{gather}
    \label{eq:taylor_spp}
	v^{(i,p_i)}(q_i) = q_i^{p_i} \\
	V_{p_1\cdots p_N} = \frac{1}{p_1! \cdots p_N!}\frac{\partial^{p_1 + \cdots + p_N} V}{\partial q_1^{p_1} \cdots \partial q_N^{p_N}},
\end{gather}
and the storage requirement for the integrals of all single-particle
potentials~(\ref{eq:taylor_spp}) is $\mathcal{O}(N \bar P K^2)$ where $\bar P$
is the highest power in the expansion (commonly $\bar P=4$ or $6$).  When HO
basis functions are used, the integrals are analytic, e.g.,
\begin{equation}
\begin{split}
    \langle \chi_\mu | \bar q^2 | \chi_\nu \rangle &= 
    \sqrt{(\nu+2)(\nu+1)}\delta_{\mu,\nu+2} + (2\nu+1)\delta_{\mu\nu} \\
    &\hspace{1em}+ \sqrt{(\nu-2)(\nu-1)}\delta_{\mu,\nu-2},
\end{split}
\end{equation}
and are identical for all modes (when using $\bar{q}$ instead of $q$),
which reduces the memory requirement to only  $\mathcal{O}(\bar P K^2)$.

\begin{table}[t]
\begin{tabular*}{0.48\textwidth}{@{\extracolsep{\fill}}lcccccc}
\hline\hline
    Molecule       & $N$ & Ref.~values & $N_\mathrm{states}$ & $\epsilon_1$  & $\epsilon_2$ & $\epsilon_2^d$ \\ \hline 
    Acetonitrile   & 12 & Ref.~\onlinecite{avci-1} & 70 & 2.0           & 0.1          & 1.0            \\
    Ethylene       & 12 & Tight VHCI+SPT2          & 100 & 2.0           & 0.1          & 1.0             \\
    Ethylene Oxide & 15 & Ref.~\onlinecite{avci-1} & 200 & 5.0           & 0.1          & 1.0            \\
    Naphthalene    & 48 & Ref.~\onlinecite{schaefer-2018} & 25 & 3.0    & 0.005        & 0.75           \\ 
    \hline\hline
\end{tabular*}
\caption{Molecules studied in this work, including the dimensionality of their
potential energy surface ($N$), the reference values used for comparison, the
number of states we calculate ($N_\mathrm{states}$), and the VHCI+PT2
parameters used ($\epsilon$ in cm$^{-1}$).  For all stochastic and semistochastic PT2 calculations, we
average over $N_s = 50$ samples with $N_c = 100$ configurations per sample.
}
\label{tab:params}
\end{table}

For the sake of comparison, we study the same four molecules as in previous
work on VHCI~\cite{vhci-1}: acetonitrile, ethylene, ethylene oxide, and naphthalene.
For acetonitrile (a common benchmark), we use a quartic PES with normal modes
calculated with CCSD(T)/cc-pVTZ and anharmonic force constants with
B3LYP/cc-pVTZ.\cite{vib-rrbpm, c2h4_pes} 
For ethylene, we use a sextic PES calculated entirely using
CCSD(T)/cc-pVQZ~\cite{c2h4_pes}.  
For ethylene oxide, we use a quartic
PES with normal modes calculated with CCSD(T)/cc-pVTZ and anharmonic force
constants with
B3LYP/6-31+G(d,p)~\cite{ethylene_oxide_pes,avci-1,vasci-1,carrington-2016,schaefer-2018}.
Finally, for naphthalene, we use a PES
calculated entirely with B97-1/TZ2P~\cite{naph_pes}.  The PES parameters used
in this study can be found in the Supplementary Material.  The dimensionality,
references for benchmark values, the number of states we calculate, and the
VHCI and PT2 parameters are given in Tab.~\ref{tab:params}; these latter values
are used throughout the work except where otherwise stated.
All calculations were performed with in-house software on a single node using
12 cores and 96 GB of memory.

\subsection{Impacts of heat-bath criteria and harmonic oscillator versus VSCF modals} \label{sec:results-hovsvscf}

\begin{figure}[t]
	\includegraphics[width=\columnwidth]{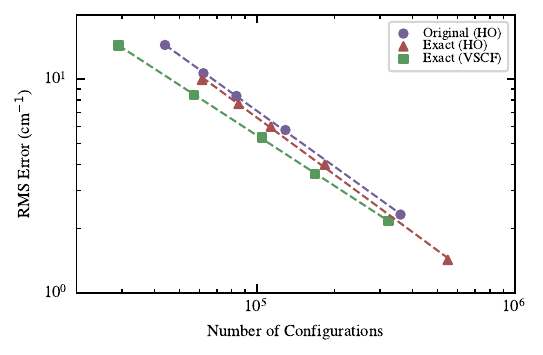}
        \caption{Performance of VHCI using the original heat-bath (HB)
criterion with HO modals (purple circles), the new, ``exact'' HB criterion with HO
modals (red triangles), and the new, ``exact'' HB criterion with VSCF modals
(green squares).  The RMS error for the first 200 states of ethylene oxide is
plotted with respect to number of configurations in the variational space
(obtained by varying $\epsilon_1$). A fit to the form $\mathrm{Error} =
a/N_\mathrm{conf}$ is also shown for each series.}
	\label{fig:conv}
\end{figure}

In Sec.~\ref{sec:theory-vhci}, a new heat-bath criterion was introduced that is
more precise than the one used previously~\cite{vhci-1}.  In
Fig.~\ref{fig:conv}, we show, for ethylene oxide, the RMS error with the previous criterion and the
new ``exact'' criterion (using HO modals) as a function of the number of
configurations included in the variational calculations (obtained by varying
$\epsilon_1$). The exact heat-bath criterion consistently improves on the
original one; however, with this many configurations, the improvement is only marginal
(less than 1~cm$^{-1}$).  Nonetheless, the added cost of the new
criterion is negligible, and so we will use it throughout the rest of the work.

In Fig.~\ref{fig:conv}, we also show the RMS error obtained with VSCF modals
(and the new exact heat-bath criterion). We see that for small configuration spaces
(near $10^4$), the use of VSCF modals improves results by about
10~cm$^{-1}$, which drops to about 1~cm$^{-1}$ for larger configuration spaces
(near $10^5$).  As expected, for larger configuration spaces, we approach the
full CI limit, and the discrepancy is reduced.  The improved performance of
VSCF-based VHCI is attributable to the mean-field treatment of anharmonicity in
the reference modals. However, our results were performed with VSCF optimized
for the ground state; when targeting a broad range of vibrational eigenstates,
one could imagine modal optimization via state-averaged
VHCI-SCF~\cite{sa-casscf-1, mcscf-2}.

\begin{figure}[t]
    \includegraphics[scale=0.95]{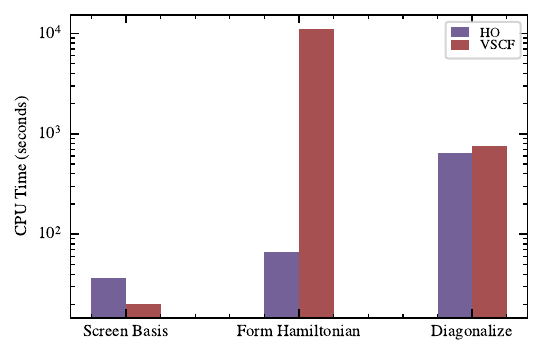}
    \caption{Time required for each part of a variational VHCI calculation of the first 200
states of ethylene oxide with $\epsilon_1 = 5.0$, using harmonic oscillator (purple) or VSCF
(red) reference modals.}
    \label{fig:eo_timing_vhci}
\label{fig:eo_time}
\end{figure}

In addition to accuracy, we must consider the computational cost associated
with HO or VSCF modals.  In Fig.~\ref{fig:eo_time} (still for ethylene oxide), 
we show timings of the various VHCI steps with $\epsilon_1 = 5.0$.  First, we
note that the time required to screen the basis of configurations is
negligible, indicating that the heat-bath approach to selection is effective at
eliminating this cost.  Next, we see that forming the Hamiltonian is
significantly more expensive with VSCF modals than with HO modals. This is
because with HO modals, the Hamiltonian matrix is far more sparse and
predictably so. With VSCF modals, the matrix is less sparse and the matrix
elements are more expensive to calculate due to the integral transformations. 
 In fact, we see that with VSCF modals, the formation of the Hamiltonian even
overwhelms the cost of the subsequent diagonalization, whose timing is similar
for HO and VSCF references because both add a similar number of configurations
to the variational space for a fixed $\epsilon_1$.  These observations apply
even more so to PT2 calculations, since a much larger number of matrix elements
need to be calculated.

\begin{figure*}[t]
\includegraphics[width=\textwidth]{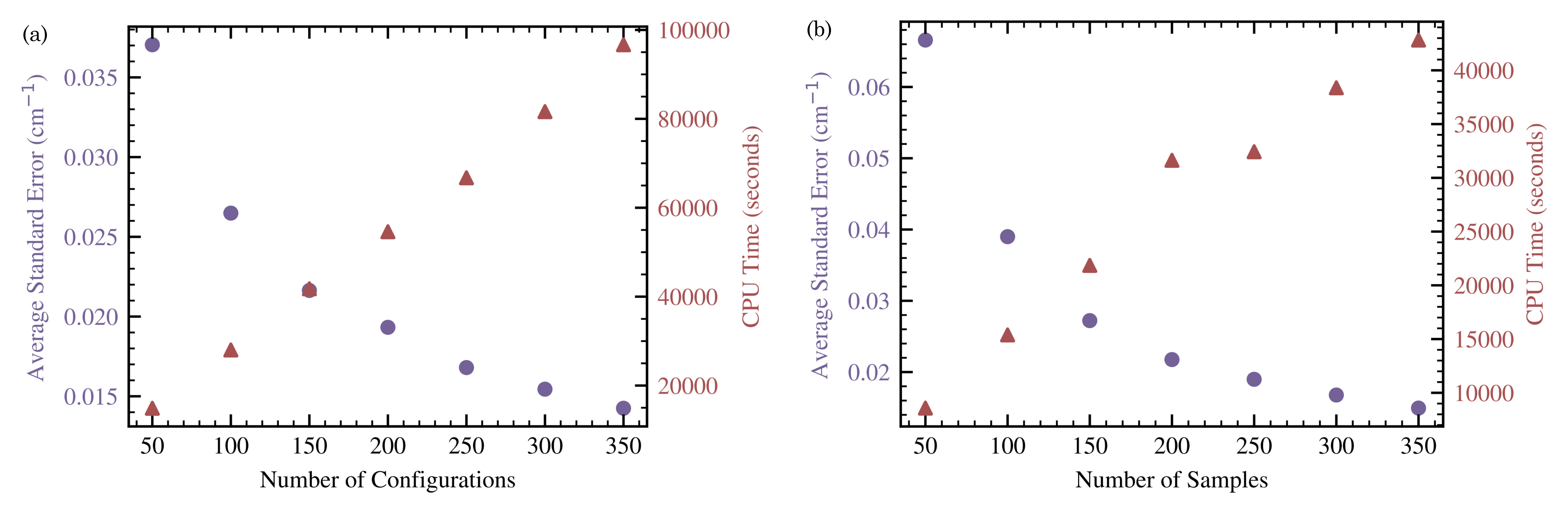}
\caption{The average standard error (purple circles) and CPU timing (red triangles) 
for acetonitrile with VHCI+SSPT2. Results are shown as a function of
the number of configurations $N_c$ with $N_s=100$ (a) and as a function
of the number of samples $N_s$ with $N_c=50$ (b).
}
\label{fig:sspt2_stats}
\end{figure*}

We conclude that the minor improvement in accuracy with VSCF modals is offset
by the significant increase in computational cost, which is entirely due to the
complexity of building the Hamiltonian matrix (for variational calculations)
and for calculating Hamiltonian matrix elements (for PT2 calculations).  For
larger molecules, where the relative number of configurations that can be
included is smaller, or for more anharmonic systems than those studied here,
perhaps the increased cost of the VSCF modals is worthwhile.  For the rest of
this work, we only use HO modals.  Results using VSCF modals can be found in
the Supplemental Information.

\subsection{VHCI with SPT2 and SSPT2}

We next assess the performance of stochastic and semistochastic perturbation
theory, which are the primary advances of the present work.  We apply these
methods to all four test molecules.
In Fig.~\ref{fig:sspt2_stats}, we demonstrate, for acetonitrile, the validity of our
semistochastic implementation.  We see the decrease in the (average) standard
error of the mean and the increase in CPU time as the number of configurations $N_c$ (a)
or the number of samples $N_s$ (b) is increased. 
As expected, standard error decreases as $N_s^{-1/2}$
and the CPU time increases linearly with $N_s$.
For these parameters, the standard errors are seen to be very small, even for
relatively small $N_s$ and $N_c$.  For the rest of the results, we use $N_c =
100$ and $N_s=50$.

\begin{figure}[b]
	\includegraphics[width=\columnwidth]{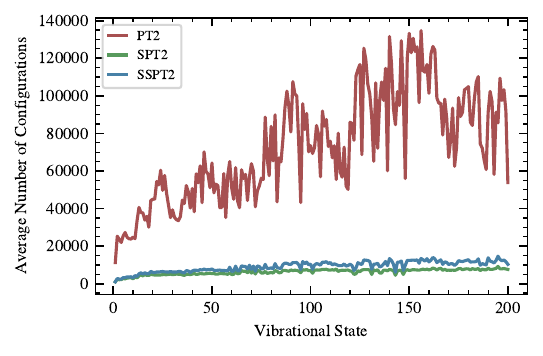}
        \caption{Average number of configurations used for the PT2
(red), SPT2 (green), and SSPT2 (blue) corrections for each
vibrational state of ethylene oxide. The variational VHCI calculation produced
about $10^5$ configurations.}
	\label{fig:eo_pt2states}
\end{figure}

We emphasize that the benefit of SPT2 and SSPT2 is a reduction in memory (the
calculation time may or may not be reduced, depending on the acceptable
statistical error).  In Fig.~\ref{fig:eo_pt2states}, we show, for ethylene
oxide, the average number of perturbing configurations included per sample, for
each state, comparing deterministic PT2, SPT2, and SSPT2.  The required memory
is directly proportional to the number of perturbing configurations per sample.
We see that the stochastic techniques reduce the number of perturbing
configurations by about a factor of 10, enabling calculations to be performed
with a smaller memory requirement (or equivalently, with a smaller value of
$\epsilon_2$ for the same amount of memory).

\begin{figure}[b]
	\includegraphics[width=\columnwidth]{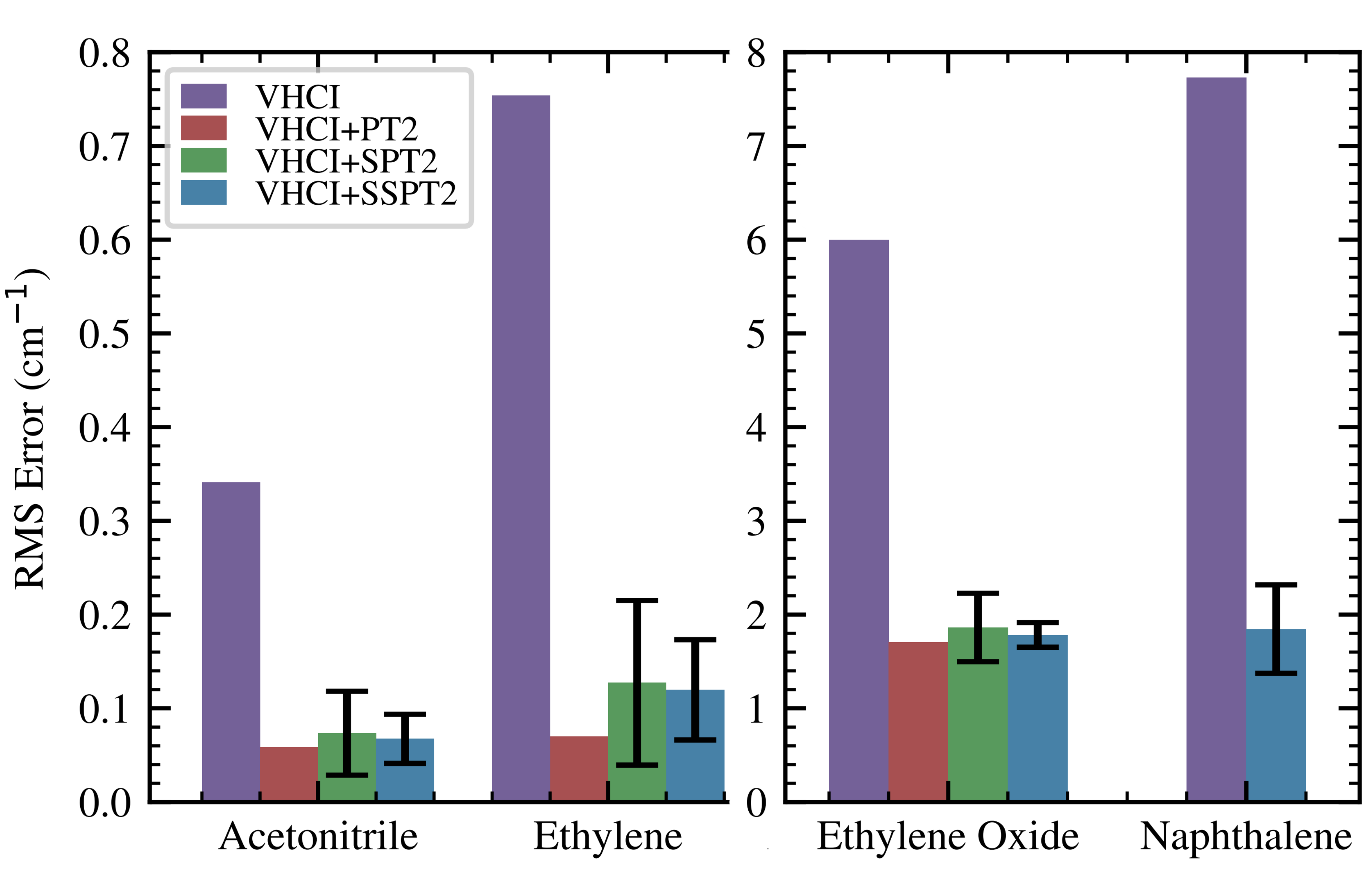}
	\caption{
RMS error for the lowest lying vibrational states of the molecules indicated
using VHCI (purple) with PT2 (red), SPT2 (green), and SSPT2 (blue)
corrections. For naphthalene, PT2 calculations are impossible with our chosen $\epsilon_2$.
The SPT2 and SSPT2 data include a $\pm 2\sigma$ error bar showing the standard error 
averaged over all states.}
	\label{fig:results}
\end{figure}

Finally, we present results for all four test molecules in
Fig.~\ref{fig:results}.  As expected, the PT2 correction significantly improves
the performance of variational VHCI. Moreover, the stochastic and
semistochastic corrections reproduce the PT2 energies within their statistical
error bars, using only a fraction of the memory. 
For the smaller molecules, acetonitrile and ethylene, we can easily achieve
0.1~cm$^{-1}$ accuracy. 
For the larger molecules, ethylene oxide and naphthalene, we can achieve
1-2~cm${-1}$ accuracy, although it's unclear whether the reference values
are converged to a precision better than this.

The power of semistochastic PT2 is best demonstrated on our largest molecule,
naphthalene, by comparing to the previous VHCI report with deterministic
PT2~\cite{vhci-1}. In that work and this work, the size of the 
variational space is about the same, containing about $2\times 10^6$ configurations.
With deterministic PT2, only $\epsilon_2 = 0.2$ was possible, which led
to a calculation time of about $4,000$ core hours. With SSPT2, we were able to
use a much smaller $\epsilon_2 = 0.005$ (impossible with deterministic PT2)
while only needing to treat about $2\times 10^5$
perturbing configurations at a time. A value $\epsilon_2^d = 0.75$ controlled the stochastic
error and the total time was only 800 core hours.

\section{Conclusion}
\label{sec:conclusion}
We have presented several advances to VHCI. 
A new ``exact'' heat-bath criterion was proposed, and VHCI was reformulated to use configurations of VSCF modals. 
The new heat-bath criterion allows for implicit sorting of exact matrix elements and improves the calculation with a negligible cost increase. 
However, for the systems studied here, we found that using VSCF modal configurations is not worth the additional cost of matrix element evaluations and integral transformations. 
Most significantly, the original memory bottleneck of the PT2 corrections has been eliminated through the use of a (semi)stochastic implementation, and stochastic errors can be easily controlled to within 1 cm$^{-1}$. 

We are confident that VHCI is capable of providing benchmark vibrational energies on model systems of the type studied here.
The next goal is to apply it to larger systems with stronger anharmonicity and more realistic representations of the PES.
We have already outlined the formalism for SOP potentials, and extension to $n$-mode potentials is an important future aim.
For very large molecules, a natural use for VHCI is as an active space solver, similar to its primary use in electronic structure theory.
For example, active spaces defined through localized normal modes, such as those that optimize the ground state VSCF wavefunction~\cite{ocvscf-1, ocvscf-2, ocvscf-3, ocvscf-4, ocvscf-5} or that optimize a localization cost function,\cite{local_normal_modes-1, local_normal_modes-2, local_normal_modes-3, local_normal_modes-4} will be interesting avenues for VHCI.

\section*{Supplementary material}

See the supplementary material for the potential energy surfaces of all the systems studied and extraneous results for the same systems in this work using VSCF modal references and tighter parameters.

\section*{Acknowledgements}
This work was supported by the U.S. Department of Energy, Office of Science,
Basic Energy Sciences, under Award No.~DE-SC0023002.
We acknowledge computing resources from Columbia University's
Shared Research Computing Facility project, which is supported by NIH Research
Facility Improvement Grant 1G20RR030893-01, and associated funds from the New
York State Empire State Development, Division of Science Technology and
Innovation (NYSTAR) Contract C090171, both awarded April 15, 2010.

\section*{Data availability statement}
The data that support the findings of this study are available from the
corresponding author upon reasonable request.

\raggedbottom

\end{document}


\title{Supplementary Material: Vibrational heat-bath configuration interaction with semistochastic perturbation theory 
	using harmonic oscillator or VSCF modals}

\author{Henry K. Tran and Timothy C. Berkelbach}

\maketitle
\beginsupplement

\section{Potential Energy Surfaces}
The potential energy surfaces used in the main text are presented in the supplementary information. The format is as follows. \\
\indent\texttt{Modes: n} \\
\indent\texttt{0 w\textsubscript{1}} \\
\indent\texttt{1 w\textsubscript{2}} \\
\indent$\vdots$\\
\indent\texttt{n w\textsubscript{n+1}} \\
\indent\texttt{Force\_constants: N}\\
\indent\texttt{p\textsubscript{1} q\textsubscript{1} $\cdots$ q\textsubscript{p\textsubscript{1}} F\textsubscript{1}}\\
\indent$\vdots$\\
\indent\texttt{p\textsubscript{N} q\textsubscript{1} $\cdots$ q\textsubscript{p\textsubscript{N}} F\textsubscript{N}}\\
where \texttt{w\textsubscript{i}} is the normal mode frequency in wavenumbers of the $n$ different modes. \texttt{p\textsubscript{i}} is the power of the $i$\textsuperscript{th} force constant, and every term \texttt{q\textsubscript{j}} after it are the modes for which the derivatives are taken with respect to. The value of the force constant, scaled by the square root of the normal mode frequencies, is \texttt{F\textsubscript{i}}. For example, the line \\
\indent \texttt{3 0 0 1 100} \\
indicates
\begin{equation}
	\frac{1}{\sqrt{\text{\texttt{w\textsubscript{1} w\textsubscript{1} w\textsubscript{2}}}}} \frac{\partial^3 V}{\partial^2\texttt{q\textsubscript{1}} \partial \texttt{q\textsubscript{2}}} = 100
\end{equation}
The PES of systems studied in this paper are given with the filename suffix \texttt{.inp}. 

\section{Individual State Results}
The state energies from each calculation is provided as well with format \texttt{mol\_e1\_e2.csv} where \text{mol} is the molecule, \texttt{e1} is $\epsilon_1$, and \texttt{e2} is $\epsilon_2$.

\section{Results with VSCF-VHCI} \label{sec:results-main}
In this section, VSCF-VHCI results, which were excluded from the main text, are shown.

The first 70 vibrational states of acetonitrile are compared to A-VCI results that were used as benchmarks in the first VHCI paper. The parameters used are $\epsilon_1 = 2.0$, $\epsilon_2 = 0.1$, and $\epsilon_2^d = 1.0$. All calculations involving acetonitrile are displayed in Figure \ref{fig:ac}. The root means squared (RMS) errors are displayed in Figure \ref{fig:ac_error}. The HO-VHCI and VSCF-VHCI calculations use a variational space of 23845 and 18925, respectively. The average number of configurations used for the PT2 corrections of each vibrational state is plotted in Figure \ref{fig:ac_pt2states}. The CPU timing for each part of the calculation is shown in Figure \ref{fig:ac_timing} and the CPU timing for the VHCI calculation in particular is shown in Figure \ref{fig:ac_timing_vhci}. 
\begin{figure}
\begin{subfigure}{\columnwidth}
	\includegraphics[width=0.5\columnwidth]{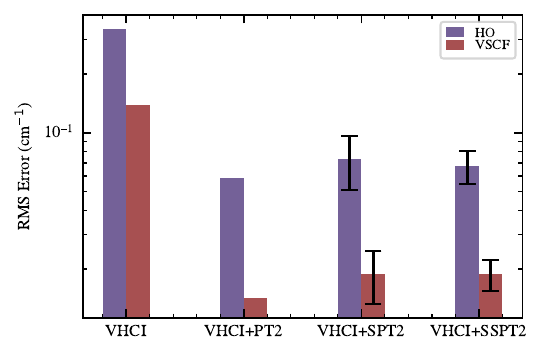}
	\caption{RMS error for the first 70 vibrational states of acetonitrile Results are compared to A-VCI calculations on the same system. The SPT2 and SSPT2 plots include an error bar indicating average standard error of means between all states. The harmonic oscillator results are plotted in orange and the VSCF results are plotted in pink.}
	\label{fig:ac_error}
\end{subfigure}
\end{figure}
\begin{figure}\ContinuedFloat
\begin{subfigure}{\columnwidth}
	\includegraphics[width=0.5\columnwidth]{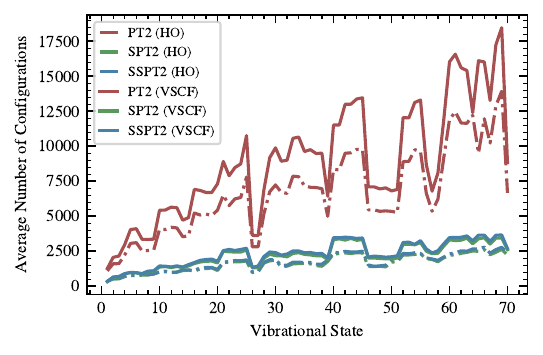}
	\caption{Average number of configurations used for the PT2, SPT2, and SSPT2 corrections for each vibrational state of acetonitrile. Different colors distinguish between PT2 (red), SPT2 (green), and SSPT2 (blue). Dotted lines are for HO-VSCF+XPT2 and full lines are for VSCF-VHCI+XPT2. HO-VHCI and VSCF-VHCI are done with 23845 and 18925 configurations, respectively.}
	\label{fig:ac_pt2states}
\end{subfigure}
\end{figure}
\begin{figure}\ContinuedFloat
\begin{subfigure}{\columnwidth}
	\includegraphics[width=0.5\columnwidth]{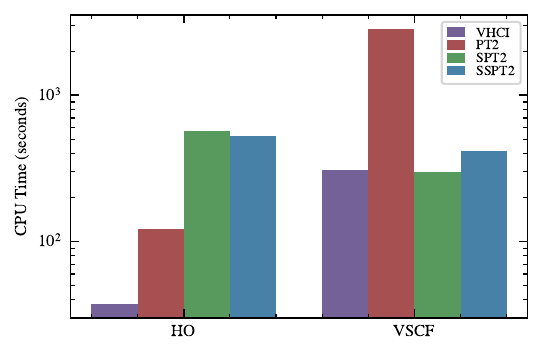}
	\caption{CPU cost in seconds for each part of the acetonitrile calculation. Different colors distinguish between VHCI (purple), PT2 (red), SPT2 (green), and SSPT2 (blue).}
	\label{fig:ac_timing}
\end{subfigure}
\end{figure}
\begin{figure}\ContinuedFloat
\begin{subfigure}{\columnwidth}
	\includegraphics[width=0.5\columnwidth]{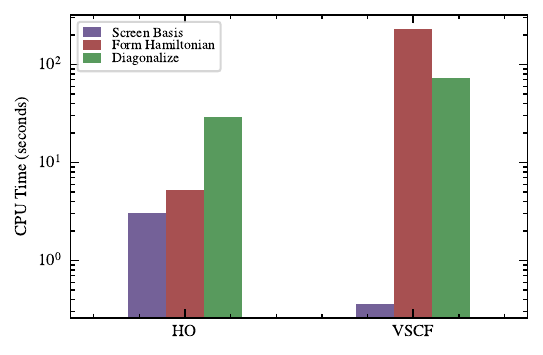}
	\caption{CPU cost in seconds for each part of the acetonitrile VHCI calculation. Screen Basis (purple) is the cost of the heat-bath criterion. Form Hamiltonian (red) is the cost of determining each matrix element and filling in the matrix. Diagonalize (green) is the cost of diagonalizing the Hamiltonian in the variational space at each VHCI iteration.}
	\label{fig:ac_timing_vhci}
\end{subfigure}
\caption{Calculation of the first 70 vibrational states of acetonitrile using a PES in literature. The parameters used are $\epsilon_1 = 2.0$, $\epsilon_2 = 0.1$, and $\epsilon_2^d=0.1$.}
\label{fig:ac}
\end{figure}

The first 200 vibrational states of ethylene oxide are compared to A-VCI results that were used as benchmarks in the first VHCI paper. The parameters used are $\epsilon_1 = 5.0$, $\epsilon_2 = 0.1$, and $\epsilon_2^d=1.0$. All calculations involving ethylene oxide are displayed in Figure \ref{fig:eo}. The root means squared (RMS) errors are displayed in Figure \ref{fig:eo_error}. The HO-VHCI and VSCF-VHCI calculations use a variational space of 113272 and 104873, respectively. The average number of configurations used for the PT2 corrections of each vibrational state is plotted in Figure \ref{fig:eo_pt2states}. The CPU timing for each part of the calculation is shown in Figure \ref{fig:eo_timing} and the CPU timing for the VHCI calculation in particular is shown in Figure \ref{fig:eo_timing_vhci}. 
\begin{figure}
\begin{subfigure}{\columnwidth}
	\includegraphics[width=0.5\columnwidth]{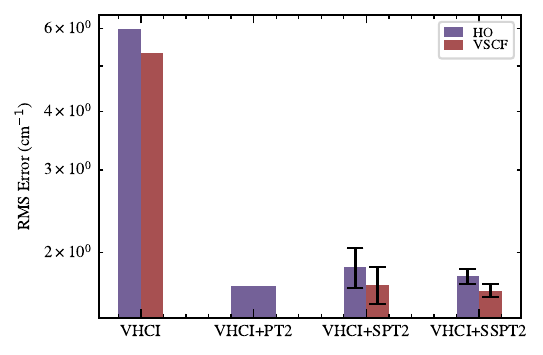}
	\caption{RMS error for the first 200 vibrational states of ethylene oxide. The SPT2 and SSPT2 plots include an error bar indicating average standard error of means between all states. The harmonic oscillator results are plotted in orange and the VSCF results are plotted in pink. Note that there is no result for VHCI+PT2 using VSCF modals.}
	\label{fig:eo_error}
\end{subfigure}
\end{figure}
\begin{figure}\ContinuedFloat
\begin{subfigure}{\columnwidth}
	\includegraphics[width=0.5\columnwidth]{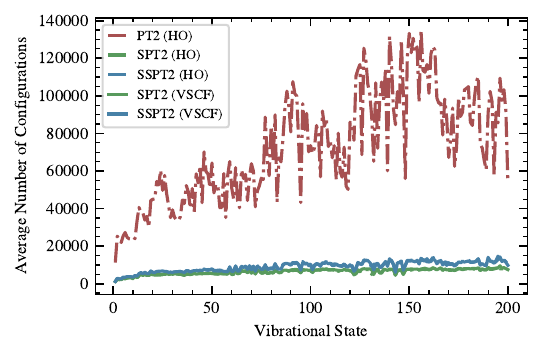}
	\caption{Average number of configurations used for the PT2, SPT2, and SSPT2 corrections for each vibrational state of ethylene oxide. Different colors distinguish between PT2 (red), SPT2 (green), and SSPT2 (blue). Dotted lines are for HO-VSCF+XPT2 and full lines are for VSCF-VHCI+XPT2. HO-VHCI and VSCF-VHCI are done with 113272 and 104873 configurations, respectively.}
	\label{fig:eo_pt2states}
\end{subfigure}
\end{figure}
\begin{figure}\ContinuedFloat
\begin{subfigure}{\columnwidth}
	\includegraphics[width=0.5\columnwidth]{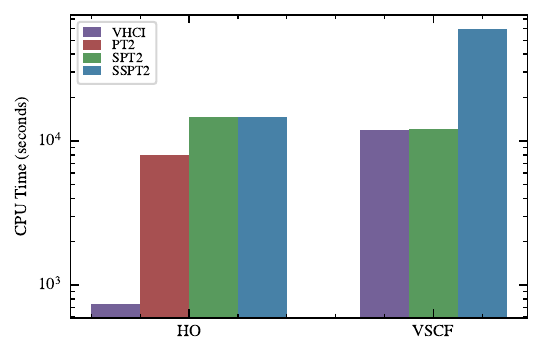}
	\caption{CPU cost in seconds for each part of the ethylene oxide calculation. Different colors distinguish between VHCI (purple), PT2 (red), SPT2 (green), and SSPT2 (blue).}
	\label{fig:eo_timing}
\end{subfigure}
\end{figure}
\begin{figure}\ContinuedFloat
\begin{subfigure}{\columnwidth}
	\includegraphics[width=0.5\columnwidth]{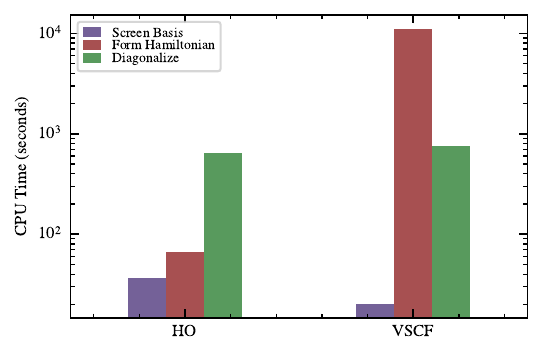}
	\caption{CPU cost in seconds for each part of the ethylene oxide VHCI calculation. Screen Basis (purple) is the cost of the heat-bath criterion. Form Hamiltonian (red) is the cost of determining each matrix element and filling in the matrix. Diagonalize (green) is the cost of diagonalizing the Hamiltonian in the variational space at each VHCI iteration.}
	\label{fig:eo_timing_vhci}
\end{subfigure}
\caption{Calculation of the first 200 vibrational states of ethylene oxide using a PES in literature. The parameters used are $\epsilon_1 = 5.0$, $\epsilon_2 = 0.1$, and $\epsilon_2^d=1.0$.}
\label{fig:eo}
\end{figure}

The first 100 vibrational states of ethylene are compared to VSCF-VHCI+SPT2 results using a tighter set of parameters ($\epsilon_1 = 0.5$ and $\epsilon_ 2 = 0.01$). All calculations involving ethylene are displayed in Figure \ref{fig:c2h4}. The parameters used are $\epsilon_1 = 2.0$, $\epsilon_2 = 0.1$ and $\epsilon_2^d=1.0$. The root means squared (RMS) errors are displayed in Figure \ref{fig:c2h4_error}. The HO-VHCI and VSCF-VHCI calculations use a variational space of 117649 and 113175, respectively. The average number of configurations used for the PT2 corrections of each vibrational state is plotted in Figure \ref{fig:c2h4_pt2states}. The CPU timing for each part of the calculation is shown in Figure \ref{fig:c2h4_timing} and the CPU timing for the VHCI calculation in particular is shown in Figure \ref{fig:c2h4_timing_vhci}. 
\begin{figure}
\begin{subfigure}{\columnwidth}
	\includegraphics[width=0.5\columnwidth]{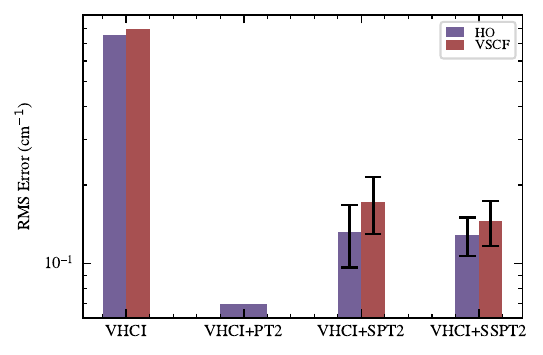}
	\caption{RMS error for the first 100 vibrational states of ethylene. Results are compared to our own calculations using VHCI+SPT2 starting from a VSCF modal basis with parameters $\epsilon_1 = 0.50$ and $\epsilon_2 = 0.01$. The SPT2 and SSPT2 plots include an error bar indicating average standard error of means between all states. The harmonic oscillator results are plotted in orange and the VSCF results are plotted in pink. Note that there is no result for VHCI+PT2 using VSCF modals.}
	\label{fig:c2h4_error}
\end{subfigure}
\end{figure}
\begin{figure}\ContinuedFloat
\begin{subfigure}{\columnwidth}
	\includegraphics[width=0.5\columnwidth]{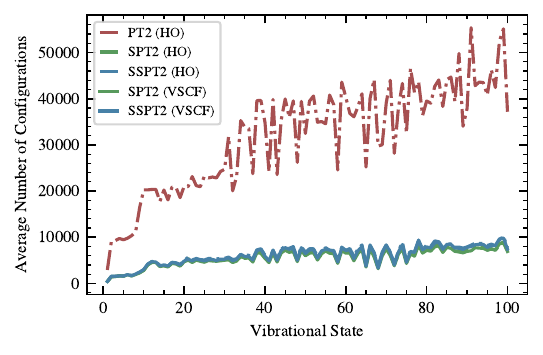}
	\caption{Average number of configurations used for the PT2, SPT2, and SSPT2 corrections for each vibrational state of ethylene. Different colors distinguish between PT2 (red), SPT2 (green), and SSPT2 (blue). Dotted lines are for HO-VSCF+XPT2 and full lines are for VSCF-VHCI+XPT2. HO-VHCI and VSCF-VHCI are done with 117649 and 113175 configurations, respectively.}
	\label{fig:c2h4_pt2states}
\end{subfigure}
\end{figure}
\begin{figure}\ContinuedFloat
\begin{subfigure}{\columnwidth}
	\includegraphics[width=0.5\columnwidth]{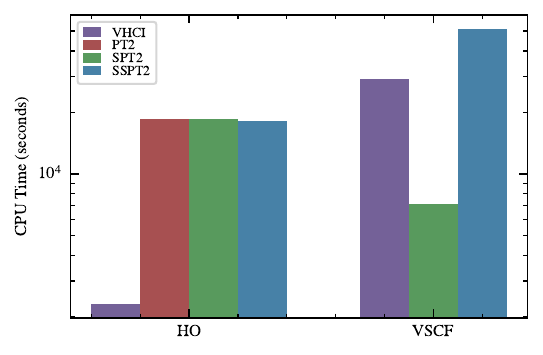}
	\caption{CPU cost in seconds for each part of the ethylene calculation. Different colors distinguish between VHCI (purple), PT2 (red), SPT2 (green), and SSPT2 (blue).}
	\label{fig:c2h4_timing}
\end{subfigure}
\end{figure}
\begin{figure}\ContinuedFloat
\begin{subfigure}{\columnwidth}
	\includegraphics[width=0.5\columnwidth]{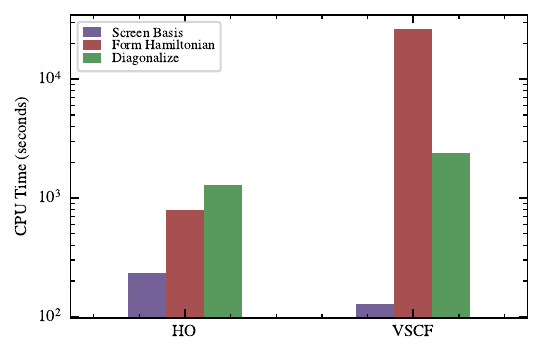}
	\caption{CPU cost in seconds for each part of the ethylene VHCI calculation. Screen Basis (purple) is the cost of the heat-bath criterion. Form Hamiltonian (red) is the cost of determining each matrix element and filling in the matrix. Diagonalize (green) is the cost of diagonalizing the Hamiltonian in the variational space at each VHCI iteration.}
	\label{fig:c2h4_timing_vhci}
\end{subfigure}
\caption{Calculation of the first 100 vibrational states of ethylene using a PES in literature. The parameters used are $\epsilon_1 = 2.0$, $\epsilon_2 = 0.1$ and $\epsilon_2^d=1.0$.}
\label{fig:c2h4}
\end{figure}

While VSCF-VHCI seems to improve over HO-VHCI generally, this is not always the case. It is interesting to note that the difference between the perturbative spaces of HO-VHCI+XPT2 and VSCF-VHCI+XPT2 is not much, and in fact, VSCF usually leads to a larger perturbative space. Despite this, the variational space is smaller for VSCF and the total space is consistently smaller for VSCF. However, the difference is still relatively small. For ethylene, the size of the total space shrinks by only $\sim$3\% between harmonic oscillators and VSCF.

The timing data is seen in Figures \ref{fig:ac_timing}, \ref{fig:eo_timing}, and \ref{fig:c2h4_timing}. First, since a relatively looser set of parameters are used for these comparisons, the immediate timing advantage is not apparent. While the SPT2 and SSPT2 calculations use a smaller perturbative space, the advantage is outweighed through the need to run multiple samples. Figure \ref{fig:tight_timing} shows the timing results using a smaller set of parameters where the advantages are more prominent. In these results, the smaller perturbative space clearly outweighs the need for multiple samples. Note that this is the total CPU time and each sample is embarrassingly parallel.
\begin{figure}
	\begin{subfigure}{\columnwidth}
	\includegraphics[width=0.5\columnwidth]{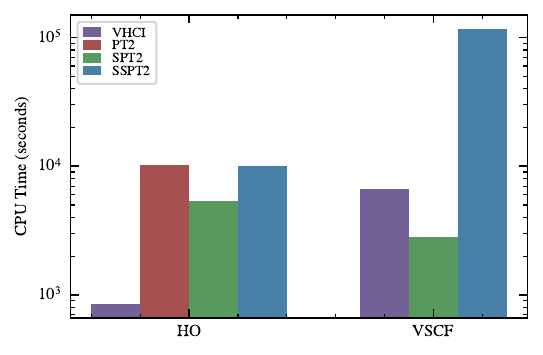}
	\caption{CPU cost in seconds for each part of the acetonitrile calculation using parameters $\epsilon_1=0.100$, $\epsilon_2 = 0.001$, and $\epsilon_2^d=0.010$. Different colors distinguish between VHCI (purple), PT2 (red), SPT2 (green), and SSPT2 (blue).}
	\label{fig:ac_tight_timing}
	\end{subfigure}
\end{figure}
\begin{figure}\ContinuedFloat
	\begin{subfigure}{\columnwidth}
		\includegraphics[width=0.5\columnwidth]{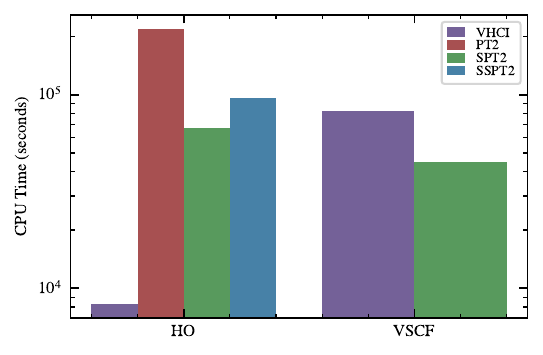}
		\caption{CPU cost in seconds for each part of the ethylene oxide calculation using parameters $\epsilon_1=1.00$, $\epsilon_2 = 0.01$, and $\epsilon_2^d=0.10$. Different colors distinguish between VHCI (purple), PT2 (red), SPT2 (green), and SSPT2 (blue).}
		\label{fig:eo_tight_timing}
	\end{subfigure}
\end{figure}
\begin{figure}\ContinuedFloat
	\begin{subfigure}{\columnwidth}
	\includegraphics[width=0.5\columnwidth]{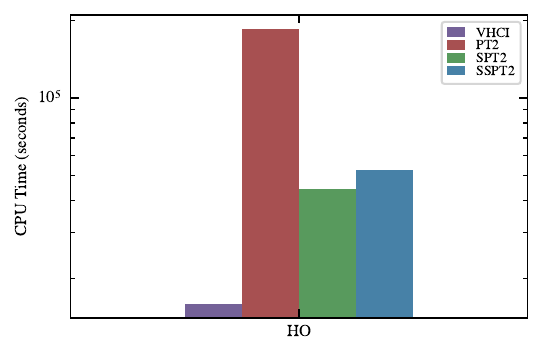}
	\caption{CPU cost in seconds for each part of the ethylene calculation using parameters $\epsilon_1=0.50$, $\epsilon_2 = 0.01$, and $\epsilon_2^d=0.10$. Different colors distinguish between VHCI (purple), PT2 (red), SPT2 (green), and SSPT2 (blue).}
	\label{fig:c2h4_tight_timing}
\end{subfigure}
\caption{CPU cost in seconds for the VHCI, VHCI+PT2, VHCI+SPT2, and VHCI+SSPT2 calculations of the systems of interest, but with smaller parameters and larger configuration spaces.}
\label{fig:tight_timing}
\end{figure}

The benefit of SPT2 is especially seen for VSCF-VHCI+S(S)PT2. Figure \ref{fig:ac_timing} has data for all methods, and VSCF-VHCI+PT2 is by far the most expensive method. It is generally too expensive to be applied to larger systems. The main difficulty is in generating the matrix elements from a VSCF reference. Harmonic oscillator reference matrix elements are much sparser, and thus the matrix is easier to generate. This can be seen in Figures \ref{fig:ac_timing_vhci}, \ref{fig:eo_timing_vhci}, and \ref{fig:c2h4_timing_vhci}. The generation of matrix elements is much more expensive for VSCF references, but the rest of the VHCI calculation is similar across harmonic oscillator and VSCF references. Hence, with SPT2 and SSPT2, the smaller perturbative space is a huge advantage for VSCF-VHCI+SPT2 and VSCF-VHCI+SSPT2 because fewer matrix elements have to be generated. This brings the computational cost of HO-VHCI+SPT2 on par with VSCF-VHCI+SPT2. VSCF-VHCI+SSPT2 tends to be quite expensive because of the deterministic space.

\section{Additional Results}
\begin{figure}
	\includegraphics[width=0.5\columnwidth]{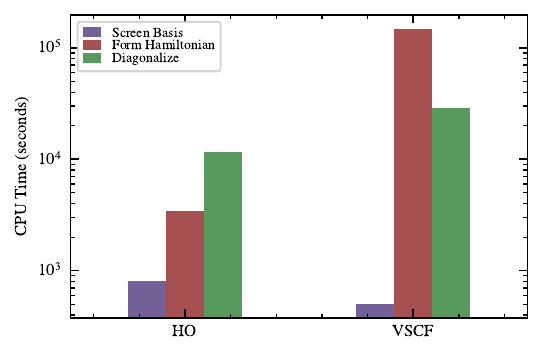}
	\caption{CPU cost in seconds for each part of the ethylene VHCI calculation with $\epsilon = 0.5$. Screen Basis (purple) is the cost of the heat-bath criterion. Form Hamiltonian (red) is the cost of determining each matrix element and filling in the matrix. Diagonalize (green) is the cost of diagonalizing the Hamiltonian in the variational space at each VHCI iteration.}
	\label{fig:c2h4_timing_vhci2}
\end{figure}

In this section, we provide calculations not presented in the main text. Most of these are calculations using a tighter set of parameters. Figure \ref{fig:ac2} shows the results for acetonitrile. The biggest difference is that it appears that HO-VHCI+SSPT2 might outperform VSCF-VHCI+SSPT2, but the difference is on the order of 10$^{-3}$ cm$^{-1}$, and it is not clear that the reference A-VCI values are correct to that order either. Figure \ref{fig:c2h4_timing_vhci2} and Figure \ref{fig:eo2} display results for ethylene and ethylene oxide. The results don't change any conclusions in the main text.
\begin{figure}
	\begin{subfigure}{\columnwidth}
		\includegraphics[width=0.5\columnwidth]{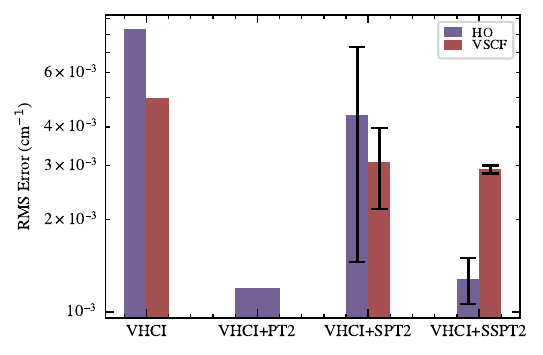}
		\caption{RMS error for the first 70 vibrational states of acetonitrile Results are compared to A-VCI calculations on the same system. The SPT2 and SSPT2 plots include an error bar indicating average standard error of means between all states. The harmonic oscillator results are plotted in orange and the VSCF results are plotted in pink.}
		\label{fig:ac_error2}
	\end{subfigure}
\end{figure}
\begin{figure}\ContinuedFloat
	\begin{subfigure}{\columnwidth}
		\includegraphics[width=0.5\columnwidth]{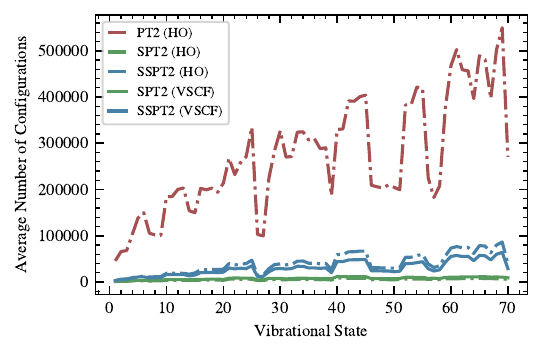}
		\caption{Average number of configurations used for the PT2, SPT2, and SSPT2 corrections for each vibrational state of acetonitrile. Different colors distinguish between PT2 (red), SPT2 (green), and SSPT2 (blue). Dotted lines are for HO-VSCF+XPT2 and full lines are for VSCF-VHCI+XPT2. HO-VHCI and VSCF-VHCI are done with 211206 and 152228 configurations, respectively.}
		\label{fig:ac_pt2states2}
	\end{subfigure}
\end{figure}
\begin{figure}\ContinuedFloat
	\begin{subfigure}{\columnwidth}
		\includegraphics[width=0.5\columnwidth]{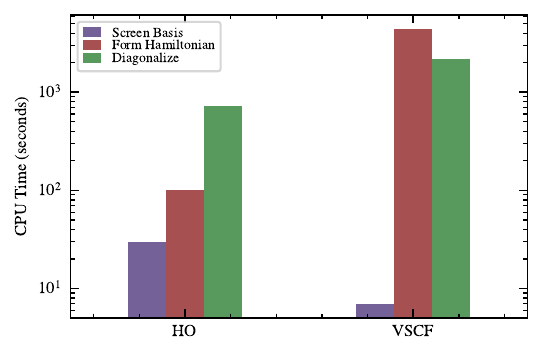}
		\caption{CPU cost in seconds for each part of the acetonitrile VHCI calculation. Screen Basis (purple) is the cost of the heat-bath criterion. Form Hamiltonian (red) is the cost of determining each matrix element and filling in the matrix. Diagonalize (green) is the cost of diagonalizing the Hamiltonian in the variational space at each VHCI iteration.}
		\label{fig:ac_timing_vhci2}
	\end{subfigure}
	\caption{Calculation of the first 70 vibrational states of acetonitrile. The parameters used are $\epsilon_1 = 0.100$, $\epsilon_2 = 0.001$, and $\epsilon_2^d=0.010$.}
	\label{fig:ac2}
\end{figure}

\begin{figure}
	\begin{subfigure}{\columnwidth}
		\includegraphics[width=0.5\columnwidth]{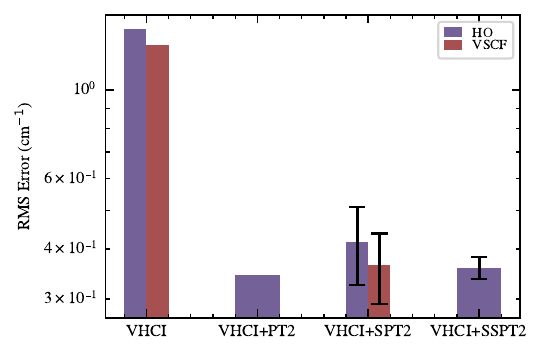}
		\caption{RMS error for the first 200 vibrational states of ethylene oxide Results are compared to A-VCI calculations on the same system. The SPT2 and SSPT2 plots include an error bar indicating average standard error of means between all states. The harmonic oscillator results are plotted in orange and the VSCF results are plotted in pink.}
		\label{fig:eo_error2}
	\end{subfigure}
\end{figure}
\begin{figure}\ContinuedFloat
	\begin{subfigure}{\columnwidth}
		\includegraphics[width=0.5\columnwidth]{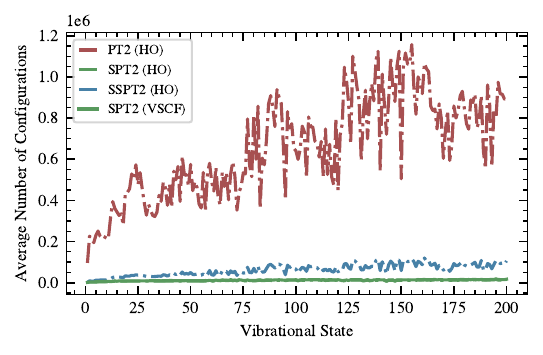}
		\caption{Average number of configurations used for the PT2, SPT2, and SSPT2 corrections for each vibrational state of ethylene oxide. Different colors distinguish between PT2 (red), SPT2 (green), and SSPT2 (blue). Dotted lines are for HO-VSCF+XPT2 and full lines are for VSCF-VHCI+XPT2. HO-VHCI and VSCF-VHCI are done with 549006 and 498764 configurations, respectively.}
		\label{fig:eo_pt2states2}
	\end{subfigure}
\end{figure}
\begin{figure}\ContinuedFloat
	\begin{subfigure}{\columnwidth}
		\includegraphics[width=0.5\columnwidth]{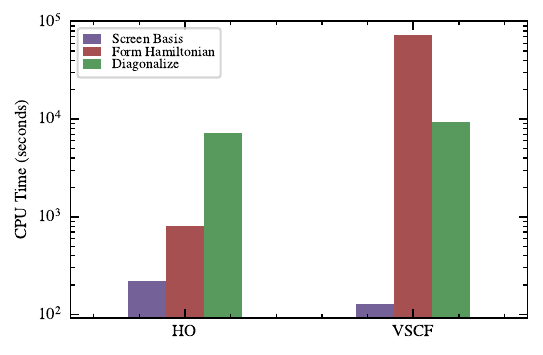}
		\caption{CPU cost in seconds for each part of the ethylene oxide VHCI calculation. Screen Basis (purple) is the cost of the heat-bath criterion. Form Hamiltonian (red) is the cost of determining each matrix element and filling in the matrix. Diagonalize (green) is the cost of diagonalizing the Hamiltonian in the variational space at each VHCI iteration.}
		\label{fig:eo_timing_vhci2}
	\end{subfigure}
	\caption{Calculation of the first 200 vibrational states of ethylene oxide. The parameters used are $\epsilon_1 = 1.00$, $\epsilon_2 = 0.01$, and $\epsilon_2^d=0.10$.}
	\label{fig:eo2}
\end{figure}

\section{SPT2 Statistics}

The main text analyzed the statistics of SSPT2 for ethylene oxide. The same analyses for SPT2 are shown in Figure \ref{fig:spt2_stats}.
\begin{figure}
	\begin{subfigure}{\columnwidth}
		\includegraphics[width=0.5\columnwidth]{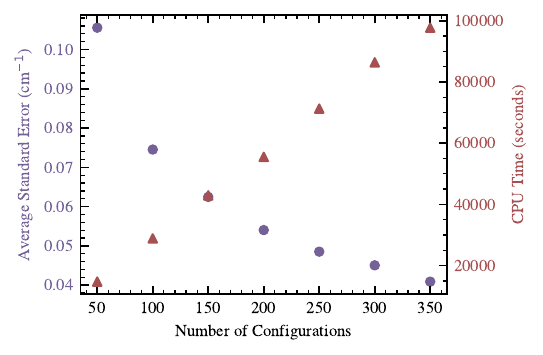}
		\caption{The average standard error (purple circle) and CPU timing (red triangle) for the first 200 states of ethylene oxide with respect to $N_s$ using HO-VHCI+SPT2.}
		\label{fig:spt2_ns_stats}
	\end{subfigure}
\end{figure}
\begin{figure}\ContinuedFloat
	\begin{subfigure}{\columnwidth}
		\includegraphics[width=0.5\columnwidth]{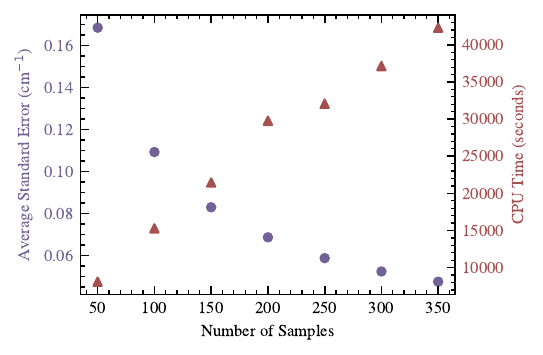}
		\caption{The average standard error (purple circle) and CPU timing (red triangle) for the first 200 states of ethylene oxide with respect to $N_c$ using HO-VHCI+SPT2.}
		\label{fig:spt2_nw_stats}
	\end{subfigure}
	\caption{The average standard error (purple circle) and CPU timing (red triangle) for the first 200 states of ethylene oxide with respect to $N_s$ or $N_c$ using HO-VHCI+SPT2. The parameters for the ethylene oxide calculation are $\epsilon_1=2.0$ and $\epsilon_2 = 0.1$.}
	\label{fig:spt2_stats}
\end{figure}